\documentstyle[12pt,epsfig]{article}
\newcommand{\case}[2]{\mbox{\footnotesize $\displaystyle \frac{#1}{#2}$}}

\begin{document}
\begin{center}
{\Large \bf  Heavy Meson Observables and \\Dyson-Schwinger Equations}
\vskip 0.3 true in
{\large M. A. Ivanov,\footnotemark[1]
Yu. L. Kalinovsky,\footnotemark[2]
P. Maris$\,$\footnotemark[3]
and C. D. Roberts$\,$\footnotemark[4]}
\date{}
\vskip 0.3 true in
{\it 
\footnotemark[1]Bogoliubov Laboratory of Theoretical Physics, \\
Joint Institute for Nuclear Research, 141980 Dubna,
Russia\vspace*{0.2\baselineskip}\\ 
\footnotemark[2]Laboratory of Computing Techniques and Automation, \\
Joint Institute for Nuclear Research, 141980 Dubna,
Russia\vspace*{0.2\baselineskip}\\ 
\footnotemark[3]Center for Nuclear Research, Physics Department,
Kent State University,\\ Kent Ohio 44242, USA\vspace*{0.2\baselineskip}\\
\footnotemark[4]Physics Division, Bldg. 203, Argonne National Laboratory, \\
Argonne Illinois 60439, USA }
\end{center}

\begin{abstract}%
Dyson-Schwinger equation (DSE) studies show that the $b$-quark mass-function
is approximately constant, and that this is true to a lesser extent for the
$c$-quark.  These observations provide the basis for a study of the leptonic
and semileptonic decays of heavy pseudoscalar mesons using a ``heavy-quark''
limit of the DSEs.  When exact, this limit reduces the number of independent
form factors.  Semileptonic decays with light mesons in the final state are
also accessible because the DSEs provide a description of light-quark
propagation characteristics and light-meson structure.  A description of
$B$-meson decays is straightforward, however, the study of decays involving
the $D$-meson indicates that $c$-quark mass-corrections are quantitatively
important.
\end{abstract}

Dyson-Schwinger equations provide a nonperturbative, Poincar\'e invariant,
continuum approach to studying quantum field theories: two familiar examples
are the gap equation in superconductivity and the Bethe-Salpeter equation
describing relativistic 2-body bound states.  As a system of coupled integral
equations, a truncation of the DSEs is necessary to obtain a tractable
problem.  The simplest truncation scheme is a weak-coupling expansion, which
generates every diagram in perturbation theory.  Hence, in the intelligent
application of DSEs to QCD, there is a tight constraint on the ultraviolet
behaviour of the Schwinger functions (dressed-propagators and vertices).
That is crucial in extrapolating into the infrared, in constructing uniformly
valid symmetry-preserving truncations, and in developing phenomenological
models necessary for anticipating the results of the current generation of
hadron physics facilities.

The development of efficacious truncations is not a purely algebraic task,
and neither is it always obviously systematic.  Nevertheless, it has become
clear$\,$\cite{bender96} that truncations which preserve the global
symmetries of a theory; for example, chiral symmetry in QCD, are relatively
easy to define and implement and, while it is more difficult to preserve
local gauge symmetries, much progress has been made with Abelian
theories$\,$\cite{ayse97} and more is being learnt about non-Abelian ones.
In addition, contemporary phenomenological applications now address a wide
range of observables$\,$\cite{cdranu}, yielding qualitatively robust results
and a much-needed intuitive understanding of many observables inaccessible in
perturbation theory.

A salient feature of the phenomenological application of DSEs is the
significant role played by the necessary momentum-dependent modification of
gluon and quark propagators: they are modified in perturbation theory and
this modification persists and grows in the nonperturbative domain.  For
example, in a general covariant gauge the dressed-gluon propagator is
characterised by a single scalar function, which we denote ${\cal D}(k^2)$.
Many studies of the gluon DSE show that ${\cal D}(k^2)$ is strongly enhanced
in the infrared; i.e, its behaviour in the vicinity of $k^2=0$ can be
represented as a distribution~\cite{bp89}, while for $k^2>1$-$2\,$GeV$^2$ the
perturbative result is reliable.  With such behaviour manifest in the
quark-quark interaction, dynamical chiral symmetry breaking (DCSB) and
confinement follow {\it without} fine-tuning$\,$\cite{fred}.

\begin{figure}[t]
\vspace*{-1.0em}

\centering{\ 
\epsfig{figure=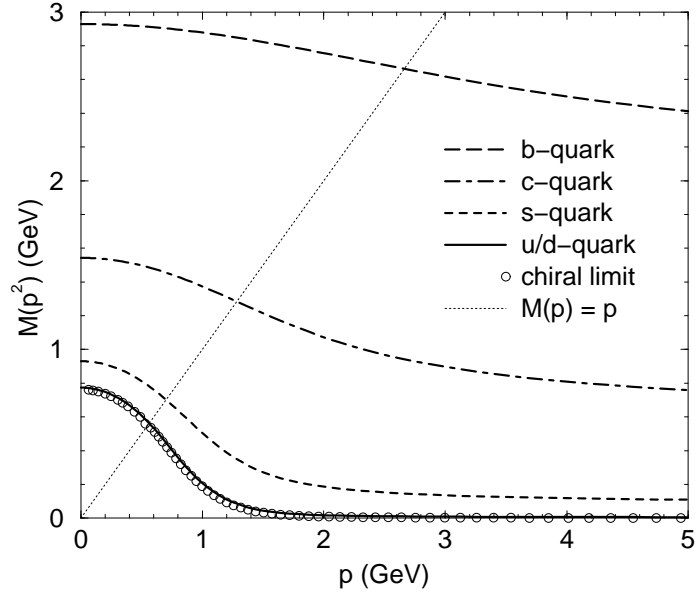,height=9.0cm}}\vspace*{-2.0em}
\caption{$M(p^2):= B(p^2)/A(p^2)$  obtained in solving the quark DSE.  The
solution of $M^2(p^2)=p^2$ defines $M^E$, the Euclidean constituent-quark
mass.
\label{plotMpp}}\vspace*{-1.0em}
\end{figure}

These phenomena can be addressed through the DSE for the dressed-quark
propagator:
\begin{equation}
\label{Sp}
S(p)  :=  \frac{1}{i\gamma\cdot p + \Sigma(p)}
 =  \frac{1}{i\gamma\cdot p\,A(p^2) + B(p^2)} 
 = i\gamma\cdot p \,\sigma_V(p^2) + \sigma_S(p^2) \,,
\end{equation}
where $\Sigma(p)$ is the renormalised dressed-quark self energy, which satisfies
\begin{equation}
\label{gendse}
\Sigma(p)  =  ( Z_2 -1)\, i\gamma\cdot p + Z_4\,m^\zeta
+\, Z_1\, \int^\Lambda_q \,
g^2 D_{\mu\nu}(p-q) \frac{\lambda^a}{2}\gamma_\mu S(q)
\Gamma^a_\nu(q,p), 
\end{equation}
with $\Gamma^a_\nu(q;p)$ the dressed-quark-gluon vertex, $m^\zeta$ the
current-quark mass, $Z_1$, $Z_2$ and $Z_4$ renormalisation constants, and
$\zeta$ the renormalisation point.  $\int^\Lambda_q := \int^\Lambda d^4
q/(2\pi)^4$ represents mnemonically a {\em translationally-invariant}
regularisation of the integral, with $\Lambda$ the regularisation mass-scale.
With the infrared-enhanced interaction introduced in Ref.~\cite{mr97} and
current-quark masses corresponding to
\begin{equation}
\label{monegev}
\begin{array}{llll}
m_{u,d}^{1\,{\rm GeV}} &
m_s^{1\,{\rm GeV}}     & 
m_c^{1\,{\rm GeV}}     & 
m_b^{1\,{\rm GeV}}     \\
 6.6\, {\rm MeV} & 
 140\,{\rm MeV}  & 
 1.0\,{\rm GeV}  & 
 3.4\,{\rm GeV}
\end{array}
\end{equation}
one obtains$\,$\cite{mr98} the dressed-quark mass function depicted in
Fig.$\,$\ref{plotMpp}.
It is clear that for light quarks ($u$, $d$ and $s$) there are two distinct
domains: perturbative and nonperturbative.  In the perturbative domain the
magnitude of $M(p^2)$ is governed by the the current-quark mass, while for
$p^2< 1\,$GeV$^2$ the mass-function rises sharply.  This is the
nonperturbative domain where the magnitude of $M(p^2)$ is determined by the
DCSB mechanism; i.e., the enhancement in the dressed-gluon propagator.

For a given flavour, the ratio ${\cal L}_f:=M^E_f/m_f^\zeta$ is a single,
quantitative measure of the importance of the DCSB mechanism in modifying
that quark's propagation characteristics.  As illustrated in
Eq.$\,$(\ref{Mmratio}),
\begin{equation}
\label{Mmratio}
\begin{array}{l|c|c|c|c|c}
\mbox{\sf flavour} 
        &   u,d  &   s   &  c  &  b  &  t \\\hline
 \frac{M^E}{m^{\zeta\sim 20\,{\rm GeV}}}
       &  150   &    10      &  2.3 &  1.4 & \to 1
\end{array}
\end{equation}
this ratio provides a natural classification of quarks as either light or
heavy.  For light-quarks ${\cal L}_f$ is characteristically $10$-$100$ while
for heavy-quarks it is only $1$-$2$.  The values of ${\cal L}_f$ signal the
existence of a characteristic DCSB mass-scale: $M_\chi$. At $p^2>0$ the
propagation characteristics of a flavour with $m_f^\zeta< M_\chi$ are
significantly altered by the DCSB mechanism, while for flavours with
$m_f^\zeta\gg M_\chi$ it is irrelevant, and explicit chiral symmetry breaking
dominates.  It is apparent that $M_\chi \sim 0.2\,$GeV$\,\sim \Lambda_{\rm
QCD}$.

This forms the basis for a simplification of the study of heavy-meson
observables~\cite{prc} that we summarise herein.  It motivates an
exploration of the fidelity of the approximation
\begin{equation}
\label{sb}
S_{c,b}(p) = \frac{1}{i \gamma\cdot p + \hat M_{c,b}}\,,
\end{equation}
where $\hat M_{c,b} \sim M^E_{c,b}$, so that with 
$p_{\mu} := m_{H}\,v_\mu := (\hat M_{f_Q} + E)\, v_\mu\,,$
the heavy-quark propagator is 
\begin{equation}
\label{hqf}
S_{c,b}(k+p) = \case{1}{2}\,\frac{1 - i \gamma\cdot v}{k\cdot v - E}
+ {\rm O}\left(\frac{|k|}{\hat M_{c,b}},
                \frac{E}{\hat M_{c,b}}\right)\,.
\end{equation}
($v_\mu$ is the heavy meson velocity, $v^2=-1$, and $E>0$ is the difference
between the heavy-meson mass and the effective-mass of the heavy-quark.)
Many simplifications follow from neglecting the $1/\hat M$-corrections; e.g.,
it reduces the number of independent form factors required to describe
heavy-meson $\to$ heavy-meson decays, relating them to a minimal number of
so-called ``universal'' form factors, which is a characteristic feature of
``heavy-quark'' symmetry~\cite{IW90}.  The magnitude of $M_b^E$ makes it
likely that Eq.~(\ref{hqf}) is reliable for the $b$-quark, however, in
employing the same reduction for the $c$-quark, one may expect quantitatively
important corrections.

The light quark propagators are not limited in this way.  They retain their
full momentum dependence, which is efficaciously characterised in the
parametrisation$\,$\cite{brt96}
\begin{eqnarray}
\label{SSM}
\bar\sigma^f_S(x)  & =  & 
        2 \bar m_f {\cal F}(2 (x + \bar m_f^2))
        + {\cal F}(b_1 x) {\cal F}(b_3 x) 
        \left( b^f_0 + b^f_2 {\cal F}(\epsilon x)\right)\,,\\
\label{SVM}
\bar\sigma^f_V(x) & = & \frac{2 (x+\bar m_f^2) -1 
                + e^{-2 (x+\bar m_f^2)}}{2 (x+\bar m_f^2)^2}\,,
\end{eqnarray}
where: $f=u,s$ (isospin symmetry is assumed); 
${\cal F}(y):= (1-{\rm e}^{-y})/y$;
$x=p^2/(2 D)$; $\bar m_f$ = $m_f/\sqrt{2 D}$; and
\begin{eqnarray}
\bar\sigma_S^f(x)  :=  \sqrt{2 D}\,\sigma_S^f(p^2)\,,&& 
\bar\sigma_V^f(x)  :=  2 D\,\sigma_V^f(p^2)\,,
\end{eqnarray}
with $D$ a mass scale.  This algebraic form combines the effects of
confinement and dynamical chiral symmetry breaking with free-particle
(asymptotically-free) behaviour at large, spacelike-$p^2$.  The parameters:
$\bar m_f$, $b_{0\ldots 3}^f$. in Eqs.~(\protect\ref{SSM}) and
(\protect\ref{SVM}) take the values
\begin{equation}
\label{tableA} 
\begin{array}{cccccc}
        & \bar m_f& b_0^f & b_1^f & b_2^f & b_3^f \\\hline
 u:  & 0.00897 & 0.131 & 2.90 & 0.603 & 0.185 \\
 s:  & 0.224   & 0.105 & \underline{2.90} & 0.740 & \underline{0.185}
\end{array}\,,
\end{equation}
which were determined$\,$\cite{brt96} in a least-squares fit to a range of
light-hadron observables.  The values of $b_{1,3}^s$ are underlined to
indicate that the constraints $b_{1,3}^s=b_{1,3}^u$ were imposed.  The scale
parameter $D=0.160\,$GeV$^2$.

The heavy-quark expansion introduced above can be employed in the analysis of
semileptonic pseudoscalar $\to$ pseudoscalar decays:
$P_{H_1}(p_1) \to P_{H_2}(p_2)\, \ell \,\nu\,,$
where $P_{H_1}$ represents either a $B$ or $D$ meson with momentum $p_1$
$(p_1^2= -m_{H_1}^2)$ and $P_{H_2}$ can be a $D$, $K$ or $\pi$ meson with
momentum $p_2$ $(p_2^2= -m_{H_2}^2)$.  (Light $\to$ light transitions are
discussed in Ref.~\cite{kpi}.)  The invariant amplitude describing the decay
is
\begin{equation}
A(P_{H_1} \to P_{H_2}\ell\nu) = 
\frac{G_F}{\surd 2} \,V_{qQ} \,
\bar\ell \gamma_\mu (1 -\gamma_5)\nu\, M_\mu^{P_{H_1} P_{H_2}}(p_1,p_2)\,,
\end{equation}
where $G_F$ is the Fermi weak-decay constant, $V_{qQ}$ is the appropriate
element of the Cabibbo-Kobayashi-Maskawa matrix ($q$ denotes a light-quark
and $Q$ a heavy-quark) and the hadronic current is
\begin{equation}
M_\mu^{P_{H_1} P_{H_2}}(p_1,p_2)  := 
\langle P_{H_2}(p_2)| \bar q \gamma_\mu Q | P_{H_1}(p_1)\rangle
\label{fpfm}
 =  f_+(t) (p_1 + p_2)_\mu + f_-(t) q_\mu\,,
\end{equation}
with $t := - q^2:= - (p_1-p_2)^2$.  The form factors, $f_\pm(t)$, contain all
the information about strong interaction effects in these processes and their
accurate estimation is essential to the extraction of $V_{qQ}$ from a
measurement of a semileptonic decay rate.  In impulse approximation
\begin{eqnarray}
\label{ia}
\lefteqn{ M_\mu^{P_{H_1} P_{H_2}}(p_1,p_2) = }\\
&& \nonumber
\frac{N_c}{16\pi^4}\,
\int d^4k\,
{\rm tr}\left[
\bar\Gamma_{H_2}(k;-p_2) 
S_q(k+p_2) 
i \gamma_\mu
S_Q(k+p_1)
\Gamma_{H_1}(k;p_1) 
S_{q^\prime}(k)\right]\,.
\end{eqnarray}

Hitherto unspecified in Eq.~(\ref{ia}) are $\Gamma_{H_1}(k;p_1)$ and
$\Gamma_{H_2}(k;p_2)$, the meson Bethe-Salpeter amplitudes.  They can be
obtained by solving the Bethe-Salpeter equation in a truncation consistent
with that employed in the quark DSE.  However, since for light-quarks we have
parametrised that solution, we follow Ref.~\cite{brt96} and do the same for
the light-meson amplitude; i.e., for the $\pi$- and $K$-mesons we assume
$\Gamma_{\pi,K}(k;P)= i\gamma_5\,{\cal E}_{\pi,K}(k^2)$ and employ the algebraic
parametrisation~\cite{brt96}:
\begin{equation} 
{\cal E}_{\pi,K}(k^2)  =  
\frac{\surd 2}{f_{\pi,K}}\,
\frac{C_0 \,{\rm e}^{-k^2/[2 D]} + \left.\sigma_S(k^2)\right|_{m_f=0}}
        {\left.\sigma_V(k^2)\right|_{m_f=0}}\,,
\end{equation}
which in concert with Eqs.~(\ref{SSM}) and (\ref{SVM}) provides an
efficacious algebraic representation of 
$\chi_{\pi,K}(k;P):= S(q+ P/2)\,\Gamma_{\pi,K}(k;P)\, S(q-P/2)$.
$C_0= 0.214\,$GeV is chosen to yield a calculated value $f_\pi= 0.131$.
$f_K= 0.160\,$GeV.

For a heavy-meson, Bethe-Salpeter equation studies~\cite{bsesep} suggest the
{\it Ansatz}
\begin{equation}
\label{hmbsa}
\Gamma_{H_{f}}(k;p_1) = 
\gamma_5 \left(1 + \case{1}{2} i \gamma\cdot v\right)
\case{1}{{\cal N}_{H_{f}}}\,\varphi(k^2)\,,
\end{equation}
where, using Eq.~(\ref{hqf}), the canonical normalisation condition is
\begin{equation}
\label{bsanormH}
{\cal N}_{H_{f}}^2 = \frac{1}{m_{H_{f}}}\,\frac{N_c}{32 \pi^2}
\int_0^\infty du \,
\varphi(z)^2\,
\left(\sigma_S^f(z) + \sqrt{u} \,\sigma_V^f(z)\right)
:= \frac{1}{m_{H_{f}} \kappa_f^2} \,,
\end{equation}
with $z= u - 2 E \sqrt{u}$ and $f$ labelling the light-quark flavour.  In a
solution of the Bethe-Salpeter equation the form of $\varphi(k^2)$ is
completely determined.  However, here it characterises our Ansatz and we
choose
\begin{equation}
\label{phia}
\varphi(k^2) = \exp\left(-k^2/\Lambda^2\right)\,,
\end{equation}
where $\Lambda$ is a free parameter.  As long as $\varphi(k^2)$ is a
non-negative, non-increasing function of $k^2$, calculated results are
insensitive to its detailed form.  The leptonic decay constant in the
heavy-quark limit is straightforward to determine once the Bethe-Salpeter
amplitude is known:
\begin{equation}
\label{fheavy}
f_{H_f} = \frac{\kappa_f}{\surd m_{H_f}}
\frac{N_c}{8\pi^2}\,\int_0^\infty\,du\,(\surd u - E)\,
\varphi(z)\,\left[\sigma_S^f(z) + \case{1}{2}\surd u \, \sigma_V^f(z)
\right]\,,
\end{equation}
from which it is clear that 
\begin{equation}
\label{fscaling}
f_{H_f} \surd m_{H_f} =\, {\rm const.}
\end{equation}
From Eqs.~(\ref{hqf}), (\ref{hmbsa}), (\ref{bsanormH}) and (\ref{fscaling}),
and the pseudoscalar meson mass formula~\cite{mr97}:
\begin{eqnarray}
\label{massform}
f_H\,m_H^2 & = & {\cal M}_H^\zeta\,r_H^\zeta\, ,\;\;
{\cal M}_H := {\rm tr}_{\rm flavour}
\left[M_{(\zeta)}\,\left\{T^H,\left(T^H\right)^{\rm t}\right\}\right]\,,\\
i r_H^\zeta & = & Z_4\int^\Lambda \frac{d^4q}{(2\pi)^4}\,\case{1}{2} {\rm
tr}\left[\left(T^H\right)^{\rm t} \gamma_5 {\cal S}(q_+) \Gamma_H(q;P) {\cal
S}(q_-)\right]\,,
\end{eqnarray}
where $M_{(\zeta)}= {\rm diag}(m_u^\zeta,m_d^\zeta,m_s^\zeta,\ldots)$ and
$T^H$ is a flavour matrix identifying the channel under consideration, it
also follows$\,$\cite{cdranu} that
\begin{equation}
m_{H_f} \propto \hat m_Q
\end{equation}
in the heavy-quark limit, where $\hat m_Q$ is the renormalisation point
invariant current-quark mass.  The linear trajectory becomes apparent for
$m_H\geq m_K$~\cite{cdranu,mr98}.  In contrast, for small current-quark
masses, Eq.~(\ref{massform}) yields what is commonly known as the
Gell-Mann--Oakes--Renner relation.  Thus, in Eq.~(\ref{massform}) one has a
single, exact formula that provides a unified description of light- and
heavy-meson masses.

Using Eqs.~(\ref{hqf}) and (\ref{hmbsa}) one finds$\,$\cite{mishaA} from
Eqs.~(\ref{fpfm}) and (\ref{ia}) that the $B_f \to D_f$ decay is particularly
simple to study in the heavy-quark limit.  It is described by one form
factor:
\begin{eqnarray}
\label{xifa}
f_\pm(t) & = & \case{1}{2}\, 
\frac{m_{D_f} \pm m_{B_f}}{ \sqrt{m_{D_f} m_{B_f}} }\,\xi_f(w) \,,\\
\label{xif}
\xi_f(w) & = & \kappa_f^2\,\frac{N_c}{32\pi^2}\,
\int_0^1 d\tau\,\frac{1}{W}\,
\int_0^\infty du \, \varphi(z_W)^2\,
        \left[\sigma_S^f(z_W) + \sqrt{\frac{u}{W}} \sigma_V^f(z_W)\right]\,,
\end{eqnarray}
with $W= 1 + 2 \tau (1-\tau) (w-1)$, $z_W= u - 2 E \sqrt{u/W}$
and
\begin{equation}
w = \frac{m_{B_f}^2 + m_{D_f}^2 - t}{2 m_{B_f} m_{D_f}} = v_{B_f} \cdot
v_{D_f}\,.
\end{equation}
The minimum physical value of $w$ is $w_{\rm min}=1$, which corresponds to
maximum momentum transfer with the final state meson at rest; the maximum
value is $w_{\rm max} \simeq (m_{B_f}^2 + m_{D_f}^2)/(2 m_{B_f} m_{D_f}) =
1.6$, which corresponds to maximum recoil of the final state meson with the
charged lepton at rest.  The canonical normalisation of the Bethe-Salpeter
amplitude, Eq.~(\ref{bsanormH}), automatically ensures that
\begin{equation}
\xi_f(w=1) = 1\,.
\end{equation}

Equation~(\ref{xifa}) illustrates a general result: in the heavy-quark limit,
the semileptonic decays of heavy mesons are described by a single, universal
function: $\xi_f(w)$.

The analysis of heavy $\to$ light decays is more difficult because, as
remarked above, the current-quark mass of the $u$- and $s$-quarks $m_{u/s}
\leq M_\chi \sim O(\Lambda_{\rm QCD})$.  Hence the momentum-dependent
modification of the dressed-quark propagator cannot be ignored, and the
description of these decays requires a good understanding of light-quark
propagation characteristics and the internal structure of light-mesons.  The
form factor that determines the width is
\begin{equation}
\label{fphl}
f_+^{H_1 H_2}(t) = \kappa_{q^\prime}
         \frac{\surd 2}{f_{H_2}}\,\frac{N_c}{32\pi^2}\,
        F_{q^\prime}(t;E,m_{H_1},m_{H_2}) \,,
\end{equation}
where
\begin{equation}
F_{q^\prime}(t;E,m_{H_1},m_{H_2}) =
\frac{4}{\pi}\,\int_{-1}^{1}\,\frac{d\gamma}{\sqrt{1-\gamma^2}}\,
        \int_0^1\,d\nu\,
        \int_0^\infty u^2 du\,\varphi(z_1)\,
        {\cal E}(z_1)\,W_{q^\prime}(\gamma,\nu,u) \,,
\end{equation}
$z_1= u^2-2 u \nu E$, with $W_{q^\prime}(\gamma,\nu,u)$ depending on the
light-quark propagator and its derivatives$\,$\cite{prc}.

\begin{table}[t]
\begin{center}
\begin{tabular}{|l|l|l|l|}\hline
 & DATA/ESTIMATES &    $f_B=0.170\,$GeV \\\hline
$(E,\Lambda)\,$(GeV)  & & (0.442,1.408)  \\\hline
$\Sigma^2/N $        & &  0.48        \\\hline\hline
 $f_+^{B\pi}(14.9\,{\rm GeV}^2)$  &  $0.82 \pm 0.17$~\protect\cite{latt}& 
        0.84$^\dagger$ \\\hline
 $f_+^{B\pi}(17.9\,{\rm GeV}^2)$   &  $1.19 \pm 0.28$~\protect\cite{latt} & 
        1.02$^\dagger$ \\\hline
 $f_+^{B\pi}(20.9\,{\rm GeV}^2)$ &  $1.89 \pm 0.53$~\protect\cite{latt}  & 
        1.30$^\dagger$  \\\hline
Br$({B}^0\to\pi^- \ell^+\nu)$  &  
  $[1.8\pm 0.4 \pm 0.3\pm 0.2 ]\times 10^{-4}$~\protect\cite{cleo96} & 
  2.0 $\times 10^{-4}$$^\dagger$   \\ \hline
 $f_+^{B\pi}(0)$  &  $0.18 \to 0.49$~\protect\cite{lellouch96}& 0.46 \\\hline
 $f_+^{DK}(0)$  &  0.74 $\pm$ 0.03~\protect\cite{pdg96}&  $0.62$ \\\hline
$\rho^2 $    & $\displaystyle \begin{array}{l}
                                0.91 \pm 0.15 \pm 0.06 \\
                                1.53 \pm 0.36 \pm 0.14
                                \end{array}$~\cite{cesr96} 
                              & 0.87  \\\hline
$f_{B_s}\,$(GeV) & $0.195 \pm 0.035$~\cite{hqlat} & 0.184 \\\hline
$f_{B_s}/f_B$ & $1.14 \pm 0.08 $~\cite{hqlat} & 1.083 \\\hline
$f_{D}\,$(GeV) & $0.200 \pm 0.030 $~\cite{hqlat} & 0.285  \\\hline
$f_{D_s}\,$(GeV) &  $ 0.220 \pm 0.030$~\cite{hqlat}  & 0.304  \\\hline
$f_{D_s}/f_D$ & $1.10\pm 0.06$~\cite{hqlat}& 1.066  \\\hline
\end{tabular}
\end{center}
\caption{Calculated results cf.~data (experimental or lattice simulations)
when we require $f_B=0.170\,$GeV, which is the central value estimated in
Ref.~\protect\cite{hqlat}.
Quantities marked by $^\dagger$ are used to constrain the parameters
$(E,\Lambda)$ by minimising
\mbox{$\Sigma^2 := \sum_{i=1}^{N}\,([y_i^{\rm calc}-y_i^{\rm data}]/
\sigma(y)_i^{\rm
data})^2$},
where $N$ is the number of data items used.  NB: 1) the values of $f_D$ and
$f_{D_s}$ are obtained via Eq.~(\protect\ref{fscaling}) from $f_B$ and
$f_{B_s}$, respectively, using $m_B=5.27$, $m_{B_s}=5.375$, $m_D= 1.87$ and
$m_{D_s}= 1.97\,$GeV; 2) the experimental determination of $\rho^2$ is
sensitive to the form of the fitting function, e.g., see
Ref.~\protect\cite{cesr96} and Fig.~\protect\ref{figiwfn}; 3) an analysis of
four experimental measurements of $D_s\to\mu\nu$ decays yields $f_{D_s}=0.241
\pm 0.21 \pm 0.30\,$GeV~\protect\cite{expfds}.
\label{tablea}}
\end{table}
All that is necessary for the calculation of the mesonic semileptonic heavy
$\to$ heavy and heavy $\to $ light transition form factors, and heavy-meson
leptonic decay constants is now specified.  There are two free parameters:
the binding energy, $E$, introduced after Eq.~(\ref{sb}) and the width,
$\Lambda$, of the heavy meson Bethe-Salpeter amplitude, introduced in
Eq.~(\ref{phia}).  The dressed light-quark propagators and light-meson
Bethe-Salpeter amplitudes were completely fixed in the application of this
framework to the study of $\pi$- and $K$-meson properties.  The primary goal
of this study is to determine whether, with these two heavy-quark parameters,
a description and correlation of existing heavy-meson data is possible using
the DSE framework.  Some key results are presented in Table~\ref{tablea},
which also describes how the parameters $(E,\Lambda)$ were fixed.

\begin{figure}[t]
\vspace*{-1.0em}

\centering{\
\epsfig{figure=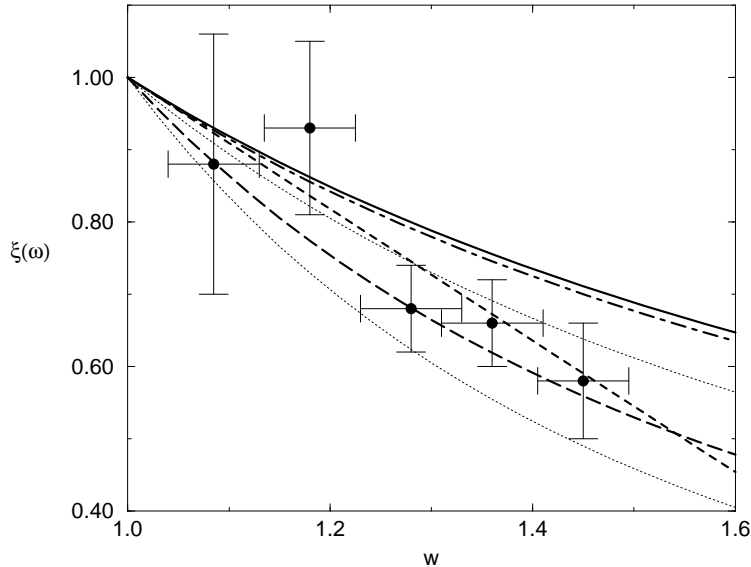,height=9.0cm}}\vspace*{-2.0em}
\caption{Calculated form of $\xi(w)$ cf. recent experimental analyses.
The solid line was obtained assuming only that the $b$-quark is heavy, the
dash-dot line assumed the same of the $c$-quark$\,$\protect\cite{prc}.
Experiment: data points - Ref.~\protect\cite{argus93}; short-dashed line -
linear fit from Ref.~\protect\cite{cesr96},
$\xi(w)  =  1 - \rho^2\,( w - 1), \; \rho^2 = 0.91\pm 0.15 \pm 0.16\,$;
long-dashed line - nonlinear fit from Ref.~\protect\cite{cesr96},
$\xi(w)  =  [2/(w+1)]\,\exp\left[(1-2\rho^2) \,(w-1)/(w+1)\right],
        \;\rho^2 = 1.53 \pm 0.36 \pm 0.14\,$.
The two light, dotted lines are this nonlinear fit evaluated with the extreme
values of $\rho^2$: upper line, $\rho^2= 1.17$ and lower line, $\rho^2=1.89$.
\label{figiwfn}}
\end{figure}
The calculated form of $\xi(w)$ is depicted in Fig.~\ref{figiwfn}.  It yields
a value of $\rho^2:= - \xi^\prime(w=1)= 0.87-0.92$,\footnote{
In this framework the minimum possible value for $\rho^2$ is
$1/3$~\protect\cite{mishaA}.}
close to that obtained with a linear fitting form$\,$\cite{cesr96}, however,
$\xi(w)$ has significant curvature and deviates quickly from that fit.  The
curvature is, in fact, very well matched to that of the nonlinear
fit$\,$\cite{cesr96}, however, the value of $\rho^2$ reported in that case is
very different from the calculated value.  The derivation of the formula for
$\xi(w)$ assumes that the heavy-quark limit, Eq.~(\ref{hqf}), is valid not
only for the $b$-quark but also for the $c$-quark.  Therefore these results
suggest that the latter assumption is only accurate to approximately 20\%;
i.e., $1/\hat M_c$-corrections are quantitatively important.

\begin{figure}[t]
\vspace*{-1.0em}

\centering{\ \epsfig{figure=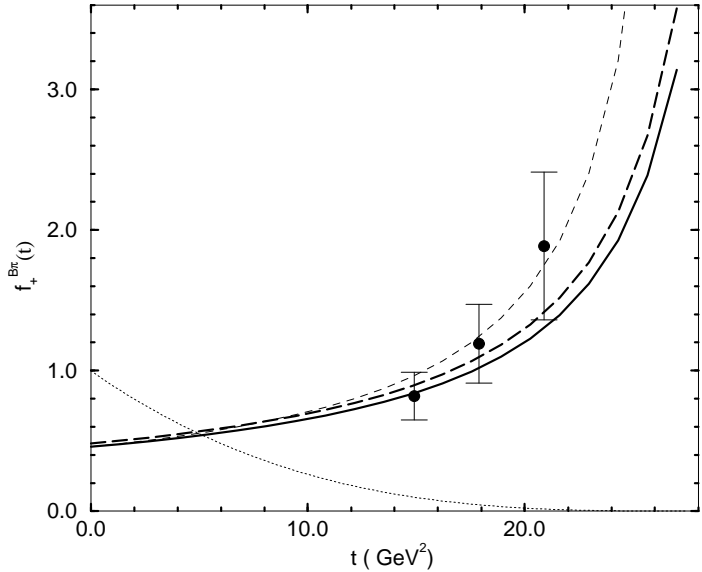,height=9.0cm}}\vspace*{-2.0em}
\caption{Calculated form of $f_+^{B\pi}(t)$.  The solid line was obtained
assuming only that the $b$-quark is heavy, the dashed line assumed the same
of the $c$-quark$\,$\protect\cite{prc}.  The data were obtained in lattice
simulation$\,$\protect\cite{latt} and the light, short-dashed line is a
vector dominance, monopole model: $f_+(t)= 0.46/(1-t/m_{B^\ast}^2)$,
$m_{B^\ast} = 5.325\,$GeV.  The light, dotted line is the phase space factor
$|f_+^{B\pi}(0)|^2 \left[(t_+-t)(t_--t)\right]^{3/2}/(\pi m_B)^3$ that
appears in the expression for the width, which illustrates that the $B\to \pi
e \nu$ branching ratio is determined primarily by the small-$t$ behaviour
$f_+^{B\pi}(t)$.  
\label{figbpi}}
\end{figure}
$f_+^{B\pi}(t)$ is depicted in in Fig.~\ref{figbpi}.  A good {\it
interpolation} of the result is provided by
\begin{equation}
f_+^{B\pi}(t)= \frac{0.458}{1 - t/m_{\rm mon}^2}\,,
\; m_{\rm mon} = 5.67\,{\rm GeV} \,.
\end{equation}
This value of $m_{\rm mon}$ can be compared with that obtained in a fit to
lattice data:$\,$\cite{latt} $m_{\rm mon}= 5.6 \pm 0.3$.  

\begin{figure}[t]
\centering{\
\epsfig{figure=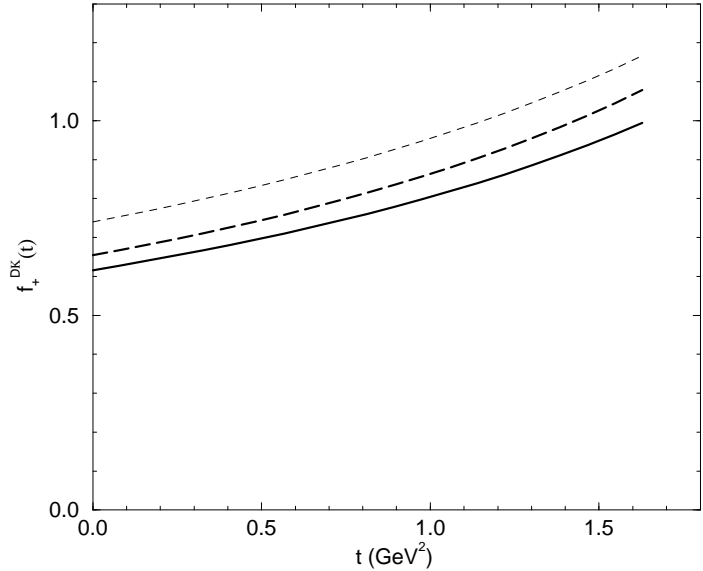,height=9.0cm}}\vspace*{-1.0em}
\caption{Calculated form of $f_+^{DK}(t)$: the solid line was obtained
assuming only that the $b$-quark is heavy, the dashed line assumed the same
of the $c$-quark$\,$\protect\cite{prc}.  The light, short-dashed line is a
vector dominance, monopole model: $f_+(q^2)= 0.74/(1-q^2/m_{D_s^\ast}^2)$,
$m_{D_s^\ast} = 2.11\,$GeV.
\label{figdk}}
\end{figure}
The calculated form of $f_+^{DK}(t)$ is depicted in Fig.~\ref{figdk}.  The
$t$-dependence is also well-approximated by a monopole fit.  The calculated
value of $f_+^{DK}(0) = 0.62$ is approximately 15\% less than the
experimental value$\,$\cite{pdg96}.  That is also a gauge of the size of
$1/\hat M_c$-corrections, which are expected to reduce the value of the $D$
and $D_s$ leptonic decay constants calculated in the heavy-quark limit: $f_D=
0.285\,$GeV, $f_{D_s}= 0.298\,$GeV.  A 15\% reduction yields $f_D =
0.24\,$GeV and $f_{D_s}=0.26\,$GeV, values which are consistent with lattice
estimates~\cite{hqlat} and the latter with experiment~\cite{expfds}.

It must be noted that Ref.~\cite{prc} explicitly {\it did not}\ assume vector
meson dominance.  The calculated results reflect only the importance and
influence of the dressed-quark and -gluon substructure of the heavy mesons.
That substructure is manifest in the dressed propagators and bound state
amplitudes, which fully determine the value of every calculated quantity.
That simple-pole {\it Ans\"atze} provide efficacious interpolations of the
calculated results on the accessible kinematic domain is not surprising,
given that the form factor must rise slowly away from its value at
\mbox{$t=0$} and the heavy meson mass provides a dominant intrinsic scale,
which is only modified slightly by the scale in the light-quark propagators
and meson bound state amplitudes.

This presentation illustrates the phenomenological application of a
heavy-quark limit of the DSEs that is based on the result that the mass
function of heavy-quarks evolves slowly with momentum.  Heavy-mesons are seen
to be little different from light-mesons: they are bound states of finite
extent with dressed-quark constituents.  The results summarised here indicate
that the heavy-quark limit can be used to develop a quantitatively reliable
description of $B$-meson observables.  However, it is inadequate for
$D$-meson observables, where corrections of 15-20\% can be expected.  A
significant feature of the DSE approach is that it provides a single
framework for the correlation of heavy $\to$ heavy and heavy $\to$ light
transitions {\it and} for their correlation with light meson observables,
which are dominated by effects such as confinement and DCSB.

This work was supported in part by the Russian Fund for Fundamental Research,
under contract 96-02-17435-a, and the US Department of Energy, Nuclear
Physics Division, under contract no. W-31-109-ENG-38.

\end{document}